\documentclass[11pt]{article}

\usepackage[margin=1in]{geometry}
\usepackage{rpmacros}
\usepackage[T1]{fontenc}
\usepackage[usenames,dvipsnames]{color}
\usepackage{stmaryrd}
\usepackage{lineno}
\usepackage{xfrac}
\usepackage{mathpazo}
\usepackage{todonotes}
\usepackage{setspace}
\usepackage{nicefrac}
\usepackage{gitinfo}
\usepackage{float}
\usepackage{scrextend}
\allowdisplaybreaks

\usepackage{fullpage}
\usepackage[utf8]{inputenc}

\usepackage{color}
\usepackage[
pagebackref, 
pdfstartview=FitH,pdfpagemode=UseNone,colorlinks=true,citecolor=blue,linkcolor=blue]{hyperref}
\hypersetup{
  linkcolor=[rgb]{0.3,0.3,0.6},
  citecolor=[rgb]{0.2, 0.6, 0.2},
  urlcolor=[rgb]{0.6, 0.2, 0.2}
}

\usepackage[nameinlink,capitalise]{cleveref}
\usepackage{amsthm}
\usepackage{thmtools,thm-restate}

\numberwithin{equation}{section}
\declaretheoremstyle[bodyfont=\it,qed=\qedsymbol]{noproofstyle}

\declaretheorem[numberlike=equation,name=Observation, Refname={Observation,Observations}]{observation}

\declaretheorem[name=Observation,numbered=no]{observation*}

\declaretheorem[numberlike=equation]{theorem}

\declaretheorem[name=Theorem,numbered=no]{theorem*}

\declaretheorem[name=Lemma,numbered=no]{lemma*}

\declaretheorem[name=Corollary,numbered=no]{corollary*}

\declaretheorem[name=Proposition,numbered=no]{proposition*}

\declaretheorem[name=Claim,numbered=no]{claim*}

\declaretheorem[name=Conjecture,numbered=no]{conjecture*}

\declaretheorem[name=Question,numbered=no]{question*}

\declaretheoremstyle[bodyfont=\it,qed=$\lrcorner$]{defstyle} 

\declaretheorem[numberlike=equation,style=defstyle]{definition}
\declaretheorem[unnumbered,name=Definition,style=defstyle]{definition*}

\declaretheorem[unnumbered,name=Example,style=defstyle]{example*}

\declaretheorem[unnumbered,name=Notation=defstyle]{notation*}

\declaretheorem[unnumbered,name=Construction,style=defstyle]{construction*}

\declaretheorem[unnumbered,name=Remark,style=defstyle]{remark*}

\usetikzlibrary{decorations.pathreplacing,arrows}

\newif\ifnote
\notefalse
\ifnote
\newcommand{\LPnote}[1]{\textcolor{BrickRed}{\guillemotleft PH: #1\guillemotright}}
\newcommand{\MKnote}[1]{\textcolor{Purple}{\guillemotleft MK: #1\guillemotright}}
\newcommand{\MSnote}[1]{\textcolor{OliveGreen}{\guillemotleft MS: #1\guillemotright}}
\newcommand{\LPnote}[1]{\textcolor{NavyBlue}{\guillemotleft SB: #1\guillemotright}}
\else
\newcommand{\LPnote}[1]{}
\newcommand{\MKnote}[1]{}
\newcommand{\MSnote}[1]{}
\fi

\newcommand{\calC}{\mathcal{C}}
\newcommand{\calL}{\mathcal{L}}
\newcommand{\calT}{\mathcal{T}}
\newcommand{\calG}{\mathcal{G}}
\DeclareMathOperator{\Diag}{Diag}
\DeclareMathOperator{\ord}{ord}

\let\epsilon\varepsilon

\onehalfspace

\title{Bivariate Linear Operator Codes
}
\author{Aaron (Louie) Putterman\thanks{School of Engineering and Applied Sciences, Harvard University, Cambridge, Massachusetts, USA. Supported in part by the Simons Investigator Awards of Madhu Sudan and Salil Vadhan, NSF Award CCF 2152413, and an HRT PhD Research Scholarship.}
\and
{ 
{Vadim Zaripov\thanks{Harvard College, Cambridge, Massachusetts, USA. Supported in part by the Harvard College Research Program.}}
}
}

\date{}

\begin{document}

\pdfoutput=1

\maketitle

\begin{abstract}
    In this work, we present a generalization of the linear operator family of codes that captures more codes that achieve list decoding capacity. Linear operator ($LO$) codes were introduced by Bhandari, Harsha, Kumar, and Sudan \cite{BHKS24} as a way to capture capacity-achieving codes. In their framework, a code is specified by a collection of linear operators that are applied to a message polynomial and then evaluated at a specified set of evaluation points. We generalize this idea in a way that can be applied to bivariate message polynomials, getting what we call \textit{bivariate linear operator} (\textit{B-LO}) codes.

    We show that bivariate linear operator codes capture more capacity-achieving codes, including permuted product codes introduced by Berman, Shany, and Tamo \cite{BST24}. These codes work with bivariate message polynomials, which is why our generalization is necessary to capture them as a part of the linear operator framework.

    Similarly to the initial paper on linear operator codes, we present sufficient conditions for a bivariate linear operator code to be list decodable. Using this characterization, we are able to derive the theorem characterizing list-decodability of $LO$ codes as a specific case of our theorem for \textit{B-LO} codes. We also apply this theorem to show that permuted product codes are list decodable up to capacity, thereby unifying this result with those of known list-decodable $LO$ codes, including Folded Reed-Solomon, Multiplicity, and Affine Folded Reed-Solomon codes.
\end{abstract}

\section{Introduction}

Error-correcting codes are used to reliably transmit information across noisy channels by encoding a message $m \in \Sigma^k$ into a codeword $c \in \Sigma^n$ in a way that $m$ can be recovered from $c$ even if some fixed fraction of coordinates of $c$ is changed. Typically, a code is represented by a set of all possible codewords $C \subseteq \Sigma^n$. Any code has two key parameters: rate, which shows how little the length of the message grows after encoding ($R = \log_{|\Sigma|}|C|/n$), and minimum (Hamming) distance, which shows how close any two codewords can be to each other ($d = \min_{c_1, c_2 \in C} |\{i \in \{0, 1, ..., n-1\} : (c_1)_i \neq (c_2)_i\}|$). Often we work with normalized distance $\delta = d/n$. The distance determines how many errors can be corrected: it is known that if the normalized distance is $\delta$, then we can uniquely reconstruct the message if the received word has errors in less than a $\delta/2$ fraction of the coordinates.

A good code reaches the optimal tradeoff between rate and distance. The \textit{Singleton bound} implies that for any code $\delta \leq 1 - R$, which means that we cannot uniquely decode any code of rate $R$ from errors in more than $(1 - R)/2$ fraction of coordinates. The Singleton bound is achieved by Reed-Solomon (RS) codes, which are obtained by evaluating message polynomials of degree less than $k$ on a fixed set of $n$ points: $C_{\text{RS}} = \{(p(a_0), p(a_1), ..., p(a_{n-1})): p \in \F[X], \deg p < k\}$. The rate of this code is $k/n$ and the normalized distance is $1 - k/n$.

We can go beyond the Singleton bound if we sacrifice the uniqueness of decoding. Elias \cite{peter1957listdecoding} suggested a notion of list decoding, where our goal is not to find the initial message uniquely but to find a small ($\text{poly}(n)$ or even constant-size) list of all potential candidates for the message. It turns out that we can list decode from a significantly higher number of errors compared to unique decoding. In particular, there are codes that are list decodable from errors in a $1 - R$ fraction of the coordinates; but for more than $1 - R$ errors the list is known to be exponential. Therefore, $1 - R$ is called \textit{list decoding capacity}.

There are several explicit code constructions that achieve list decoding capacity. The first one was introduced by Guruswami and Rudra \cite{GR08} and is called \textit{Folded Reed-Solomon} (FRS) codes. It is obtained by evaluating a message polynomial on $sn$ carefully chosen points and then bundling them into $n$ vectors of dimension $s$ each to be the encoding. Therefore, the alphabet in this case is $\F^s$, where $\F$ is the field from which we take evaluation points and polynomial coefficients.

Note that for $s = 1$ an FRS code just turns into a regular RS code, but picking $s$ to be large enough we can get a capacity-achieving code. Guruswami and Wang \cite{GW12} presented the first linear-algebraic algorithm for list-decoding FRS, while recently Kopparty \textit{et. al.} \cite{KRSW23} developed an algorithm that achieves constant list size. Therefore, while not changing either the rate or the normalized distance, this bundling operation enables the code to work well with list decoding. Moreover, most of the explicit codes that reach list decoding capacity that we know (including Multiplicity codes), use some sort of similar "bundling".

The only thing the bundling does is that it excludes some patterns of errors: now errors can only appear in "blocks" -- either the whole vector of $s$ evaluations is correct, or it is an error. For example, for $s = 5$, even if we allow $20\%$ of coordinates to be errors, it is still impossible to make every fifth evaluation an error, since in this case there would be errors in each vector, not in $20\%$ of them.

However, it is still not well known why bundling is so useful for list decoding. One attempt to explore this connection was in the work of Bhandari, Harsha, Kumar, and Sudan \cite{BHKS24}, who introduced a new family of codes called \textit{Linear Operator} ($LO$) codes, that captures many list decodable codes that use bundling. $LO$ codes use \textit{families of linear operators} of the form $\calL = (L_0, L_1, ..., L_{s-1})$, where each $L_i: \F[X] \to \F[X]$ is a linear operator and for each $p \in \F[X]$, the tuple $(L_0(p), L_1(p), ..., L_{s-1}(p))$ is denoted simply as $\calL(p)$. $LO_k^A(\calL)$, where $A\subseteq\F$ is a set of evaluation points, encodes polynomials of degree less than $k$ the following way (see \cref{def: linear op codes} for precise definition):
\begin{align*}
\F[X]_{<k} &\longrightarrow \left(\F^s\right)^n\\
p(X) &\longmapsto \left(\calL(p)(a_i)\right)_{i = 0}^{n-1}
\end{align*}

\cite{BHKS24} also introduce a notion of linearly extendible linear operator ($LELO$) codes: if for $\calL$ there exists a matrix $M(X) \in \F[X]^{s\times s}$ such that for any $p \in \F[X]$, $\calL(X\cdot p(X)) = M(X)\cdot \calL(p)$, then the corresponding code is called linearly extendible linear operator code and is labeled as $LELO_k^A(\calL)$. Many well-known codes, including RS, FRS, Multiplicity, and others, are all examples of \textit{LELO} codes. Moreover, \cite{BHKS24} presented a sufficient condition that makes an instance of a $LO$ code list decodable up to capacity, and used this criterion to show the list-decodability of these known codes:
\begin{theorem}[{\cite[Theorem 1.1]{BHKS24}}]\label{thm:bhks_intro}
Suppose $\F$ is a field of size $q$ and $L:\F[X]\to \F[X]$ a degree-preserving linear operator and $A$ a set of
	evaluation points such that for $\calL= (L^0, L^1,
	\dots,L^{s-1})$ the corresponding code $\calC$ is a linearly-extendible linear
	operator code. Furthermore, if the matrix $\Diag(\calL) \in \F^{s
		\times k}$ formed by stacking the diagonals of the $s$ linear operators
	as the rows is the generator matrix of a code with  distance
	$1-\frac{\ell}{k}$, then, $\calC$ is code with rate $\frac{k}{sn}$ and relative distance $1 -
\frac{k-1}{sn}$ over an alphabet of size $q^s$, and it is list-decodable up to the distance $1-
	\frac{k}{(s-w+1)n} - \frac1w$ with list size $q^\ell$ for any $1
	\leq w \leq s$.
\end{theorem}
$LELO$ codes greatly improved our understanding of what makes a code list decodable.

Recently, there have also been improvements in finding explicit subcodes of RS codes that are list-decodable up to capacity. In particular, Berman, Shany, and Tamo \cite{BST24} introduced a new family called \textit{Permuted Product Codes} ($PPC$). This code uses \emph{bivariate} message polynomials and is specified by two affine polynomials of coprime order $\ell_1, \ell_2$ and points $\alpha, \beta$ such that $\ell_1(\alpha) \neq \alpha$ and $\ell_2(\beta)\neq \beta$. The encoding of a polynomial $p\in \F[X, Y]$ is formed by stacking the evaluations $p(\ell_1^0(\alpha), \ell_2^0(\beta)), p(\ell_1^1(\alpha), \ell_2^1(\beta)), ..., p(\ell_1^{N-1}(\alpha), \ell_2^{N-1}(\beta))$ in columns of size $\ord(\ell_1)$. \cite{BST24} showed that $PPC$ is a subcode of RS code and that it achieves list decoding capacity. Even though this code also uses some form of bundling, the $LO$ code framework still fails to show its list decodability because of its use of bivariate message polynomials, which allows messages to have a more complex structure. Thus, the abstraction presented in \cite{BHKS24} is somehow limited and cannot capture all list decodable codes that use bundling.

The main goal of this paper is to extend the framework of \cite{BHKS24} to capture more list decodable codes, including PPCs, which then might potentially lead to discovering more capacity-achieving codes.

\subsection{Our Results}

We present \textit{Bivariate Linear Operator Codes} (\textit{B-LO}) -- a generalization of $LO$ codes that accepts bivariate message polynomials. In order to do that, we introduce \textit{bivariate families of linear operators} of the form $\calL = (L_0, L_1, ..., L_{s-1})$, $L_i: \F[X, Y] \to \F[X, Y]$. Then for any evaluation set $A = \{(x_0, y_0), (x_1, y_1), ..., (x_{n-1}, y_{n-1})\}\subseteq \F\times \F$ and degree parameters $t$ and $k$, the code $\textit{B-LO}_{t, k}^A(\calL)$ encodes bivariate polynomials of degree $<t$ in $X$ and $<k$ in $Y$ the following way:
\begin{align*}
\F[X, Y]_{\substack{<t\\<k}} &\longrightarrow \left(\F^s\right)^n \\
p(X, Y) &\longmapsto \left(\calL(p)(x_i, y_i)\right)_{i = 0}^{n-1}
\end{align*}

We first show that this is indeed a valid extension of the $LO$ code, meaning that any instance of an $LO$ code is also an instance of the \textit{B-LO} code. This implies that most of the explicit list decodable codes that we know, including FRS and Multiplicity codes, are examples of \textit{B-LO}. We will also generalize the notion of \textit{LELO} codes by introducing \textit{B-LELO}: bivariate linearly-extendible linear operator codes. These are formed by such families $\calL$ that there exist $M_x(X, Y), M_y(X, Y)\in \F[X, Y]^{s\times s}$, such that for all $p \in \F[X, Y]$, both $\calL(Xp) = M_x(X, Y)\cdot \calL(p)$ and $\calL(Yp) = M_y(X, Y)\cdot \calL(p)$.

Using this, we also show that \textit{B-LO} captures more codes than $LO$. In particular, we show that permuted product codes are an example of \textit{B-LO} codes. This means that the extension is indeed justified:

\begin{theorem}\label{thm:ppc_are_blelo}
    Any instance of the permuted product code, specified by affine polynomials $\ell_1, \ell_2 \in GA(q)$ of coprime orders $s$ and $n$ respectively, points $\alpha, \beta$ ($\ell_1(\alpha) \neq \alpha$, $\ell_2(\beta)\neq \beta$), and degree parameters $t, k$, is the same code as $\textit{B-LELO}_{t, k}^A(\calL)$, where the family of linear operators is $\calL = (L_0, ..., L_{s-1})$, $L_i(p) = p(\ell_1^i(X), \ell_2^i(Y))$, and the evaluation set is $A = \{(\alpha, \ell_2^{is}(\beta))\}_{i = 0}^{n-1}$.
\end{theorem}

The formal proof of this theorem is presented in \cref{sec:ppc_blelo}.

Finally, in our abstract framework, we present a sufficient condition that makes an instance of \textit{B-LELO} code list decodable up to capacity:
\begin{theorem}\label{thm:intro}
Suppose $\F$ is a field of size $q$, $L: \F[X, Y] \to \F[X, Y]$ is a degree-preserving linear operator, $A = \{(x_0, y_0), ..., (x_{n-1}, y_{n-1})\}\subseteq \F\times\F$ is a set of evaluation points, and $t, k$ are degree parameters such that $tk \leq sn$. Further suppose that $\calL = (L^0, L^1, ..., L^{s-1})$ is a linearly-extendible family of linear operators (where $L^i$ is $L$ composed with itself $i$ times). Then, for any $1 \leq w \leq s$ and any $d_1 \geq t, d_2\geq k$ such that $(d_1 - t + 1)(d_2-k+1) > n(s-w+1)/w$, if  $\calT = (L^0, ..., L^{s - w})$ forms a \textit{B-LELO} code $\textit{B-LELO}_{d_1, d_2}^A(\calT)$ of distance $D$ and $\Diag(\calL)\in \F^{s\times kt}$, formed by stacking the diagonals of $s$ linear operators as the rows is the generator matrix of a code of distance $1 - \ell/(tk)$, then $\textit{B-LELO}_{t, k}^A(\calL)$ is a code of rate $\frac{tk}{sn}$ that is list-decodable up to the distance $D$ with list size $q^\ell$.
\end{theorem}

The proof of this theorem generalizes the proof given in \cite{BHKS24}, but the fact that the message polynomials are \emph{bivariate} requires more complicated and careful counting to limit the space of possible solutions. We also note that the technique given in our proof generalizes well to the case of more than two variables.

We state theorem \cref{thm:intro} in a slightly more general form as \cref{thm:main}. We show that the analogous theorem for regular $LO$ codes, Theorem 5.2 of \cite{BHKS24}, is a special case of this theorem, i.e., if some $LO$ code satisfies Theorem 5.2. of \cite{BHKS24}, then if we view it as a \textit{B-LO} code, it will satisfy \cref{thm:main} of this paper and lead to the same result. We also apply our final theorem to permuted product codes, thereby showing that PPCs are list decodable up to capacity, a new derivation of the result of \cite{BST24}.

\subsection{Overview}

In \cref{sec:preliminaries} we introduce the required notation, recall the basic definitions of coding theory and the important codes that we will work with throughout the paper. We also introduce all necessary notation relevant to $LO$ codes and state the main result of \cite{BHKS24}. In \cref{sec:b_lo} we introduce bivariate linear operator codes. In \cref{sec:b_lo_examples} we present some examples of \textit{B-LO} codes and also show that \textit{LO} codes are \textit{B-LO}. In \cref{sec:list_decoding_b_lelo} we present the main theorem of the paper and show that it is indeed a generalization of Theorem 5.2 in \cite{BHKS24}. In \cref{sec:main_thm_proof} we present a proof of our main theorem. In \cref{sec:list_decoding_ppc} we apply this theorem to show the list-decodability of permuted product codes. Finally, in \cref{sec:open_questions} we present some open questions for further research.

\section{Notation \& Preliminaries}\label{sec:preliminaries}

We start with some notations that we will use throughout the paper:
\begin{itemize}
\item For a natural number $n$, $[n]$ denotes a set $\{0, 1, ..., n-1\}$.
\item $\F$ denotes a field, and for any prime power $q$, $\F_q$ denotes a finite field of $q$ elements.

\item $\F[X]$ is the ring of univariate polynomials with coefficients in $\F$, and for any $k \in \mathbb{N}$, $\F[X]_{<k}$ denotes the set of polynomials in $\F[X]$ with degree strictly less than $k$.

\item Similarly, $\F[X, Y]_{\substack{< k\\ < t}}$ denotes a set of bivariate polynomials in $\F[X, Y]$ with degree in $X$ strictly less than $k$ and degree in $Y$ strictly less than $t$.

\item For a multivariate polynomial
$f(X_0, X_1, . . . , X_{n-1}) \in \F[X_0, X_1, . . . , X_{n-1}]$, 
$deg_{X_i}(f)$ denotes
the degree of $f$, when viewing it as a univariate in $X_i$, with coefficients in the polynomial ring on the remaining variables over the field $\F$.

\item The affine group $\text{GA}(q)$ is a group whose underlying set is $\{ax + b : (a, b) \in \F_q^*\times \F_q\}$ and group operation is polynomial composition. For $\ell(x) \in \text{GA}(q)$ and $i \in \mathbb{N}^*$, let $\ell^i = \underbrace{\ell \circ \ell \circ \cdots \circ \ell}_{i}$ and $\ell^0(x) = x$. Also, let the order of $\ell$, denoted as $\ord(\ell)$ denote the smallest $i \in \mathbb{N}^*$ such that $\ell^i = x$.
\end{itemize}

\begin{definition}[codes, rate, distance]
Let $\Sigma$ be a finite alphabet and $n$ be a positive integer. Given $C \subseteq \Sigma^n$, define the following parameters $R_C$ and $\delta_C$:
\[ R_C :=\frac{\log_{|\Sigma|}(|C|)}{n}, \qquad \delta_C :=
    \min\limits_{\substack{x,y\in C\\ x\neq
        y}}\left\{\frac{\Delta(x,y)}{n}\right\} \]where $\Delta(x,y)=|\{i\in \{1,2,\ldots,n\}\colon x_i\neq y_i\}|$ denotes the Hamming distance between $x$ and $y$.
  Then, $C$ is said to be a \emph{code} of \emph{relative distance}
  $\delta_C$ and \emph{rate} $R_C$ with \emph{block length} $n$ over
  the alphabet $\Sigma$.

Alternatively, a code can be defined through an encoding map from a message space $\Sigma_{\text{msg}}^k$ to the space of codewords: $f: \Sigma_{\text{msg}}^k \to \Sigma^n$, such that the code $C$ is the image of the message space: $C = f\left(\Sigma_{\text{msg}}^k\right)$.

$\mathcal{C} = \{C_1, C_2, ..., C_n, ...\}$ is called a family of codes if for each $i \in \mathbb{N}$, $C_i$ is a code of block length $i$.
\end{definition}

\begin{definition}[linear codes] Let $\F_q$ be a field and let $\Sigma = \F_q^s$ for some
  positive integer $s$. We say that $C \subseteq (\Sigma)^n$ is a
  \emph{linear code} if $C$ is an $\F_q$-linear space when viewed as a
  subset of $\F_q^{sn}$.
\end{definition}

\begin{definition}[list decoding]
The code $C\subseteq \Sigma^n$ is said to be $(\rho, L)$-list decodable if for every $w \in \Sigma^n$:
\[
\left|\left\{c \in C : \Delta(c, w) \leq \rho n\right\}\right| \leq L
\]
In other words, $C$ is $(\rho, L)$-list decodable if for any received word $w$, there are at most $L$ codewords that disagree with $w$ on at most $\rho$ fraction of errors.
\end{definition}

If we limit the list size $L$ to be polynomial in $n$, then it is known that the maximum fraction of errors from which the code can be list decoded is known to be $1 - R_C$. This is called \textit{list-decoding capacity}.
\begin{definition}[list-decoding capacity]
Consider a family of codes $\mathcal{C} = \{C_1, C_2, ..., C_n, ...\}$, where $C_n$ is a code of block length $n$ and rate $R_n$ with alphabet $\Sigma_n$. Then $\mathcal{C}$ is said to be list decodable up to capacity if $\forall \epsilon > 0$, there exists $n_0$ such that $\forall n > n_0$ and all received words $w \in \Sigma^n_n$, there exists at most a polynomial number of codewords $c \in C_n$ such that $\Delta(c, w)/n \leq (1 - R_n(1 + \epsilon))$.

Further, if there exists an efficient algorithm for finding all these codewords, then, $\mathcal{C}$ is said to achieve
list-decoding capacity efficiently. 
\end{definition}

\subsection{Some common codes}
We also need to define several important codes that we will use in the paper.
\begin{definition}[Reed-Solomon code] For $k, n, q \in \mathbb{N}$, such that $0 \leq k \leq n \leq q$ and $q$ is a prime power, and a set $A = \{a_0, a_1, ..., a_{n-1}\} \subseteq \F_q$, the RS code $\text{RS}_{\F_q}(k, A)$ is defined as:
\begin{align*}
  \F_q[X]_{<k}  &\longrightarrow \F_q^n\\
  p(X) &\longmapsto   \left(p(a_i)\right)_{i = 0}^{n-1}
\end{align*}
In other words, the code consists of the evaluations of all polynomials with degree $< k$ on a given set of evaluation points $A$: $\text{RS}_{\F_q}(k, A) := \left\{\left(f(a_0), f(a_1),\cdots,f(a_{n-1}\right): f \in (\F_q)_{<k}[X]\right\}$ In case where the field we are using is obvious, we will denote $\text{RS}_{\F_q}(k, A)$ just as $\text{RS}(k, A)$. This code has $R = k/n$ and $\delta = 1 - (k+1)/n$.
\end{definition}

\begin{definition}[Folded Reed-Solomon code]\label{def:frs}
Let $s, n \in \mathbb{N}$ be such that $sn \leq q$, and $\gamma \in \F_q^*$ be such that $\gamma^{s-1} \neq 1$. Further, let $A = \{a_0, a_1, ..., a_{n-1}\} \subseteq \F_q$ be a set of evaluation points such that for all pairs $i$ and $j$, $i \neq j$, the sets $\{a_i, a_i\gamma, ..., a_i\gamma^{s-1}\}$ and $\{a_j, a_j\gamma, ..., a_j\gamma^{s-1}\}$ are disjoint. The FRS code $\text{FRS}_{\F_q, s, \gamma}(k, A)$ is defined as:
\begin{align*}
\F_q[X]_{<k} &\longrightarrow \left(\F_q^s\right)^n\\
p(X) &\longmapsto \left(p(\alpha_i), p\left(\alpha_i\gamma\right), ..., p\left(\alpha_i\gamma^{s-1}\right)\right)_{i=0}^{n-1}
\end{align*}
Notice that $\text{FRS}_{\F_q, s, \gamma}\left(k, A\right)$ is a code with block length $n$ over alphabet $\F_q^s$. $s$ is said to be the \emph{folding parameter}.
\end{definition}

\begin{definition}[Permuted Product code \cite{BST24}]\label{def:ppc} Let $\ell_1, \ell_2 \in \text{GA}(q)$ be two affine polynomials of coprime orders $s$ and $n$ respectively. Further let $\alpha, \beta \in \F_q$ such that $\ell_1(\alpha) \neq \alpha$, $\ell_2(\beta) \neq \beta$. Then, for any $t, k$, the permuted product code $\text{PPC}_{\F_q, \ell_1, \ell_2, \alpha, \beta}(t, k)$ is defined as:
\begin{align*}
\F_q[X, Y]_{\substack{<t\\<k}} &\longrightarrow \left(\F_q^s\right)^n\\
p(X, Y) &\longmapsto \left(p\left(\ell_1^{is}(\alpha), \ell_2^{is}(\beta)\right), p\left(\ell_1^{is+1}(\alpha), \ell_2^{is+1}(\beta)\right), \cdots, p\left(\ell_1^{is+s-1}(\alpha), \ell_2^{is+s-1}(\beta)\right)\right)_{i=0}^n
\end{align*}
In other words, the encoding of $p(X, Y)$ is created by stacking $sn$ evaluations $p\left(\ell_1^{0}(\alpha), \ell_2^{0}(\beta)\right)$, $p\left(\ell_1^{1}(\alpha), \ell_2^{1}(\beta)\right)$, ..., $p\left(\ell_1^{sn-1}(\alpha), \ell_2^{sn-1}(\beta)\right)$ into $n$ vectors in $\F_q^s$.

It is known that $PPC$ is a subcode of the RS code that achieves list-decoding capacity \cite{BST24}.
\end{definition}

\subsection{\textit{LO} codes}\label{sec:lo-codes}
In order to define our extension, we first need to define $LO$ codes. For that, we will first need several preliminary definitions from \cite{BHKS24}.

\begin{definition}[linear operators, {\cite[Definition 4.1]{BHKS24}}]\label{def:lin_op}
  Let $\calL = (L_0,\dots, L_{s-1})$ be a tuple of $s$ linear operators,
  where each $L_i:\F[X] \to \F[X]$ is an $\F$-linear operator over the
  field $\F$.  For any $f \in \F[X]$, it will be convenient to denote by
  $\calL(f)$ the (row) vector
  $(L_0(f), \dots, L_{s-1}(f)) \in \F[X]^s$.

  Given any such family $\calL$ and element $a \in \F$, define $I^a(\calL) = \{ p(X) \in \F[X] \mid  \calL(p)(a) = \bar{0} \}$.
If the family $\calL$ of linear operators family and the set of
  field elements $A \subseteq \F$ further satisfy the property that
  $I^a(\calL)$ is an ideal for each $a \in A$, we refer to the family
  $\calL$ as an \emph{ideal family of linear
    operators} with respect to $A$.

  In this case, since $\F[X]$ is a principal ideal domain, for each $a\in A$, $I^a(\calL) = \langle E^a(\calL)(X) \rangle$ for some
  monic polynomial $E^a(\calL)\in F[X]$. \qedhere
\end{definition}

\begin{definition}[linearly-extendible linear operators, {\cite[Definition 4.2]{BHKS24}}]\label{def: lin extendible lin ops}
The family $\calL$ of linear operators is said to be \emph{linearly-extendible} if there exists a matrix $M(X) \in \F[X]^{s\times s}$ such that for all $p \in \F[X]$ we have 
\[
\calL(X\cdot p(X)) = M(X) \cdot \calL(p(X)).
\]
\end{definition}
\begin{definition}[linear operator codes, {\cite[Definition 4.5]{BHKS24}}]\label{def: linear op codes}
Let $\calL=(L_0,\dots,L_{s-1})$ be a family of linear operators, $A = \{a_0, \dots,
a_{n-1}\} \subseteq \F$ be a set of evaluation points and $k$ a degree
parameter such that $k \leq s \cdot n$. Then the $LO$
code generated by $\calL$ and $A$, denoted by $LO^{A}_k(\calL)$ is given
as follows:
\begin{align*}
  \F[X]_{<k}  &\longrightarrow \left(\F^s\right)^n\\
  p(X) &\longmapsto   \left( \calL(p)(a_i) \right)_{i=0}^{n-1}\\
\end{align*}

If the $LO$ code $LO^{A}_k(\calL)$ further
  satisfies that $\calL$ is linearly-extendible, then the linear
  operator code is said to be a \emph{linearly-extendible linear
    operator code}, denoted by $LELO^{A}_k(\calL)$.
\end{definition}

It can be shown that several codes that are list decodable up to capacity, including RS, FRS, and Multiplicity codes, are all instances of the $LO$ code family. Moreover, there \cite{BHKS24} presents an easy criterion that shows when a LO code is list decodable up to capacity. Their main theorem uses one more important definition:

\begin{definition}[list-composing linear operators {\cite[Section 5.2]{BHKS24}}]
    Let $\calG = (G_0,\dots, G_{w-1})$ and
$\calT = (T_0,T_1,\dots,T_{r-1})$ be two families of linear operators
such that $G_i: \F[X] \to \F[X]$ and $\calT$ is a linearly-extendible
family of linear operators.
  We say that the pair $(\calT,\calG)$ \emph{list-composes} in terms
  of $\calL$ at the set of
  evaluation points $A$ if for every linear operator $G \in \calG$ and field
  element $a \in A$, there exists a linear function $h_{G,a}: \F^s \to \F^r$
  such that for every polynomial $f \in \F[X]$ we have
\[
  \calT(G(f))(a) = h_{G,a}(\calL(f)(a)).
\]
\end{definition}

Finally, we can present the main result of \cite{BHKS24}:

\begin{theorem}[{\cite[Theorem 5.2]{BHKS24}}]
\label{thm:bhks-main}
  If $LO^A_k(\calL)$ is an $LO$ code and there exists two
  families of linear operators $\calG= (G_0,\dots, G_{w-1})$ and $\calT =
  (T_0,\dots,T_{r-1})$ such that
  \begin{enumerate}
  \item $(\calT,A)$ forms a $LELO$ code $LELO^A_{k+nr/w}(\calT)$. \label{itm:general_lem_lin_ext}
    \item The pair $(\calT,\calG)$ list-composes in terms of $\calL$ at the set of evaluation
      points. \label{itm:general_lem_list_comp}
    \item $\calG$ is degree-preserving. \label{itm:general_lem_deg_pre}
      \item $\Diag(\calG) \in \F^{|\calG| \times k}$ (a matrix obtained by stacking the diagonals of $\calG$, when viewed as a $k\times k$ matrix) is the generator
        matrix of a code with distance $k-\ell$. \label{itm:general_lemma_G_dist}

    \end{enumerate}
    Then, $LO^{A}_k(\calL)$ is list-decodable up to the distance $1-
    \frac{k}{rn} - \frac1w$ with list size $q^\ell$.
  \end{theorem}

It can be checked that \cref{thm:bhks-main} implies \cref{thm:bhks_intro}. Applying \cref{thm:bhks-main} to certain instances of the $LO$ code proves their list decodability. In particular, \cite{BHKS24} uses it to show that FRS and Multiplicity codes are list decodable up to capacity. However, not all known list-decodable codes (in particular, permuted product codes) can fit in this framework. Below we propose an extension of the family of LO codes -- \textit{B-LO} codes, that can capture Permuted Product codes, and state the result analogous to \cref{thm:bhks-main} for this broader family.

\section{Bivariate Linear Operator codes}\label{sec:b_lo}
The key idea of this paper is to generalize $LO$ codes to allow using bivariate message polynomials, which would allow the message to have a more complex structure. In particular, it would allow to capture codes that work with bivariate message polynomials, including Permuted Product codes.

To do that, we first need to generalize some of the definitions from \cite{BHKS24}, presented in \cref{sec:lo-codes}.

\begin{definition}[bivatiate linear operators]\label{def:b_lin_op}
  Let $\calL = (L_0,\dots, L_{s-1})$ be a tuple of $s$ linear operators
  where each $L_i:\F[X, Y] \to \F[X, Y]$ is a $\F$-linear operator over the
  ring $\F[X, Y]$.  For any $p \in \F[X, Y]$, it will be convenient to denote by
  $\calL(p)$ the (row) vector
  $(L_0(p), \dots, L_{s-1}(p)) \in \F[X, Y]^s$. The vector of evaluations of $\calL(p)$ at a point $(x, y)$ will be denoted as $\calL(p)(x, y) \in \F^s$.

  Given any such family $\calL$ and element $(x, y) \in \F\times \F$, define $I^{(x, y)}(\calL) = \{ p(X, Y) \in \F[X, Y] \mid  \calL(p)(x, y) = \bar{0} \}$.
If the family $\calL$ of linear operators family and the set of
  field elements $A \subseteq \F\times\F$ further satisfy the property that
  $I^{(x, y)}(\calL)$ is an ideal for each $(x, y) \in A$, we refer to the family
  $\calL$ as an \emph{ideal family of linear
    operators} with respect to $A$.
\end{definition}
An important difference from \cref{def:lin_op} is that in this case, $\F[X, Y]$ is not a principal ideal domain, so we cannot express $I^{(x, y)}(\calL)$ as $\langle E^{(x, y)}(\calL)(X, Y) \rangle$. However, this will not be needed to prove our main result for Bivariate LO codes.

We will also need to generalize the notion of linear-extendibility:
\begin{definition}[bivariate linearly-extendible linear operators]\label{def:b_linear_ext}
The bivariate family of linear operators $\calL$ is said to be \emph{linearly-extendible} if there exist matrices $M_x(X, Y), M_y(X, Y) \in \F[X, Y]^{s\times s}$ such that for all $p \in \F[X, Y]$ we have 
  \begin{equation*}
  \begin{aligned}
    &\calL(X\cdot p(X, Y)) = M_x(X, Y) \cdot \calL(p(X, Y)),\\
    &\calL(Y\cdot p(X, Y)) = M_y(X, Y) \cdot \calL(p(X, Y)).
    \end{aligned}
    \end{equation*}
\end{definition}

Linearly-extendible families of linear operators have several important properties that we present below, following the Observation 4.4. of \cite{BHKS24}:
\begin{observation}\label{obs:b_lin_extendible} Suppose $\calL$ is linearly-extendible and $M_x(X, Y), M_y(X, Y)$ are the corresponding matrices from \cref{def:b_linear_ext}.
  \begin{itemize}
  \item For any $j\geq0$ we have $\calL(X^j\cdot p(X, Y))=(M_x(X, Y))^j\cdot \calL(p(X, Y))$ and $\calL(Y^j\cdot p(X, Y))=(M_y(X, Y))^j\cdot \calL(p(X, Y))$. Thus, by linearity of $\mathcal{L}$ we have that for any $q \in \F[X, Y]$:
  \[ \calL( q(X, Y) \cdot p(X, Y) ) = q\left(M_x(X, Y), M_y(X, Y)\right) \cdot \calL(p(X, Y)).\]
  
\item The family $\calL$ is
  completely specified by $\calL(1)$ and $M_x, M_y$. In other words:
  \[\calL(p(X, Y))=p(M_x(X, Y), M_y(X, Y))\cdot \calL(1)
  \]
  \item For every set $A$ of evaluation points, $\calL$ is an ideal
  family of linear operators with respect to $A$. This is because if
  at a point $(x, y)$ we have $\calL(p)(x, y)=0$, then for any $q \in \F[X, Y]$, $\calL(qp)(x, y) = q(M_x(X, Y), M_y(X, Y))(x, y)\cdot \calL(p)(x, y) = q(M_x(X, Y), M_y(X, Y))(x, y) \cdot 0 = 0$, meaning that for each $p \in I^{(x, y)}(\calL)$ and $q \in \F[X, Y]$, $pq \in I^{(x, y)}(\calL)$.
\end{itemize}
\end{observation}

These definitions allow us to define \textit{B-LO} codes:
\begin{definition}[bivariate linear operator codes]\label{def: linear op codes}
Let $\calL=(L_0,\dots,L_{s-1})$ be a bivariate family of linear operators, $A = \{(x_0, y_0), \dots,
(x_{n-1}, y_{n-1})\} \subseteq \F\times\F$ be a set of evaluation points and $t, k$ degree
parameters such that $tk \leq sn$. Then the \textit{B-LO} code generated by $\calL$ and $A$, denoted by $\textit{B-LO}^{A}_{t, k}(\calL)$ is given
as follows:
\begin{align*}
  \F[X, Y]_{\substack{<t\\ <k}}  &\longrightarrow \left(\F^s\right)^n\\
  p(X, Y) &\longmapsto   \left( \calL(p)(x_i, y_i) \right)_{i=0}^{n-1}\\
\end{align*}

If the \textit{B-LO} code $\textit{B-LO}^{A}_{t, k}(\calL)$ further
  satisfies that $\calL$ is linearly-extendible, then the linear
  operator code is said to be a \emph{bivariate linearly-extendible linear
    operator code}, denoted by $\textit{B-LELO}^{A}_{t, k}(\calL)$.

Note that the rate of $\textit{B-LO}_{t, k}^A(\calL)$ is $tk/(sn)$.
\end{definition}

An important difference of \textit{B-LELO} from $LELO$ is that \textit{B-LELO} might not be maximum distance separable (MDS). We know that $LELO$ is MDS because of the fact that $\F[X]$ is a principal ideal domain: if $\calL(p)(a) = 0$, then $p$ is divisible by $E^a(\calL)(X)$ (recall \cref{def:lin_op}), which allows us to bound the number of evaluation points where $\calL(p)$ turns to zero \cite{BHKS24}. However, in the case of \textit{B-LELO} codes, $\F[X, Y]$ is not a principal ideal domain, which means that we cannot in general get such a bound. Still, if certain conditions on an instance of \textit{B-LELO} are satisfied, it can still be list decoded from a large distance, which is later shown in \cref{thm:main}.

\section{Examples of \textit{B-LO} codes}\label{sec:b_lo_examples}
In this section, we will show that \textit{B-LO} codes are indeed an extension of \textit{LO} codes. This would automatically imply that FRS codes, Multiplicity codes, etc., are all instances of the \textit{B-LO} code as well. We will also show that \textit{LELO} codes are also \textit{B-LELO}. Finally, we will show that this generalization indeed captures more codes by showing that Permuted Product codes are also an instance of \textit{B-LELO}.
\subsection{\textit{LO} codes are \textit{B-LO}}\label{sec:lo_are_blo}
Take any code $LO_{k}^A(\mathcal{L})$. Define $\mathcal{L}' = (L_0', ..., L_{s-1}')$ such that $L_i'(p(X, Y)) = L_i(p(X, 0))$. Also define the evaluation set $A' = \{(a, 0) : a \in A\}$. Then, consider the code $\textit{B-LO}_{k, 1}^{A'}(\mathcal{L}')$. Any message polynomial $p$ for this code has degree $< 1$ in $Y$, so it is actually a polynomial in one variable $X$, which means that the message space is the same as for the initial code: $\F[X]_{<k} = \F[X, Y]_{\substack{<k\\<1}}$. Also note that for each $p \in \F[X]_{<k} = \F[X, Y]_{\substack{<k\\<1}}$ and for each $a \in A$, $\mathcal{L}(p)(a) = \mathcal{L}'(p)(a, 0)$, meaning that $\left(\calL(p)(a)\right)_{a \in A} = \left(\calL'(p)(x, y)\right)_{(x, y)\in A'}$. Therefore, $LO_{k}^A(\mathcal{L})$ is the same code as $\textit{B-LO}_{k,1}^{A'}(\mathcal{L}')$.

Moreover, when $\mathcal{L}$ is linearly-extendible, then, by \cref{def: lin extendible lin ops}, for all $p \in \F[X]$, get that $\calL(X\cdot p(X)) = M(X) \cdot \calL(p(X))$ for some matrix $M(X)$. This means that $\mathcal{L}'$ is also linearly extendible, since for any polynomial $p(X, Y) \in \F[X, Y]$: 
\begin{align*}
&\mathcal{L}'(X\cdot p(X, Y)) = \mathcal{L}(X\cdot p(X, 0)) = M(X)\cdot\calL(p(X, 0)) = M(X)\cdot \calL'(p(X, Y)) \\
&\calL'(Y\cdot p(X, Y)) = \calL(0\cdot p(X, 0)) = \calL(0) = 0 = 0\cdot \calL'(p(X, Y))
\end{align*}
Therefore, $\calL'$ is also list-extendible with $M_x(X, Y) = M(X)$ and $M_y(X, Y) = 0$, which implies that $LELO_k^A(\calL)$ is the same code as $\textit{B-LELO}_{k, 1}^{A'}(\calL')$.

This proves that \textit{B-LO} and \textit{B-LELO} are indeed valid extensions of \textit{LO} and \textit{LELO}, respectively.

\begin{observation}[FRS codes are \textit{B-LELO}]
    
To demonstrate the above point with an example, we can see that FRS codes are indeed \textit{B-LELO}, since we already know from \cite{BHKS24} that they are an instance of \textit{LELO}.

Consider $\text{FRS}_{s, \gamma}(k, A)$. Note that it is identical to $\textit{B-LELO}_{k, 1}^A(\calL)$, where $\calL' = (L_0, L_1, ..., L_{s-1})$ such that $L_i(p(X, Y)) = p(\gamma^iX, 0)$ and the evaluation set $A' = \{(a, 0): a \in A\}$. For this family, $M_x(X, Y)_{ij} = \gamma^iX\cdot \mathbb{I}[i=j]$ and $M_y(X)_{ij} = 0$. 
\end{observation}

\subsection{Permuted Product Codes are \textit{B-LELO} (Proof of \cref{thm:ppc_are_blelo})}\label{sec:ppc_blelo}
Recall that Permuted Product codes, introduced by \cite{BST24}, do not fit in the framework of \textit{LO} codes because they encode bivariate polynomials. We now give a proof of \cref{thm:ppc_are_blelo}, showing that \textit{B-LO} allows us to capture this family as well.

Take $PPC_{\ell_1, \ell_2, \alpha, \beta}(t, k)$, $\ord(\ell_1) = s$, $\ord(\ell_2) = n$. Then let $L(p(X, Y)) = p(\ell_1(X), \ell_2(Y))$ and let $\mathcal{L} = (L^0, L^1, ..., L^{s-1})$. With the evaluation set $A = \{(\alpha, \ell_2^{is}(\beta))\}_{i = 0}^{n-1}$, get that the encoding of $p(X, Y)$ is:
\begin{equation}
\begin{aligned}
\left(\calL(p)(a_i)\right)_{i = 0}^n &= \left(p\left(\ell_1^0(\alpha), \ell_2^{is}(\beta)\right), p\left(\ell_1^1(\alpha), \ell_2^{is+1}(\beta)\right), \cdots, p\left(\ell_1^{s-1}(\alpha), \ell_2^{is + s - 1}(\beta)\right)\right)_{i = 0}^n = \\ \\ &= \left(p\left(\ell_1^{is}(\alpha), \ell_2^{is}(\beta)\right), p\left(\ell_1^{is+1}(\alpha), \ell_2^{is+1}(\beta)\right), \cdots, p\left(\ell_1^{is + s-1}(\alpha), \ell_2^{is + s - 1}(\beta)\right)\right)_{i = 0}^n
\end{aligned}
\end{equation}

The last equality holds because $\ell_1$ has order $s$, meaning that $\ell_1^{j}(\alpha) = \ell_1^{is + j}(\alpha)$ for all $i, j$. Note that the last expression is exactly the encoding of $p$ for $PPC_{\ell_1, \ell_2, \alpha, \beta}(t, k)$ (see \cref{def:ppc}). Therefore, $\textit{B-LO}_{t, k}^A(\calL)$ is indeed the same code as $PPC_{\ell_1, \ell_2, \alpha, \beta}(t, k)$.

Moreover, $\calL$ is linearly-extendible, since $L_i(X\cdot p(X, Y)) = \ell_1^i(X)\cdot L_i(p(X, Y))$ and $L_i(Y\cdot p(X, Y)) = \ell_2^i(Y)\cdot L_i(p(X, Y))$, meaning that $M_x(X, Y)_{ij} = \ell_1^i(X)\cdot \mathbb{I}[i=j]$ and $M_y(X, Y)_{ij} = \ell_2^i(Y)\cdot \mathbb{I}[i=j]$. Therefore, PPC is indeed a \textit{B-LELO} code.

\section{List-decoding B-LELO}\label{sec:list_decoding_b_lelo}
In this section, we present the main theorem of this paper, which states the sufficient conditions for a \textit{B-LO} code to be list decodable.

Recall that in the statement of \cref{thm:bhks-main}, \cite{BHKS24} were using the notion of list-composing linear operators. We would also need an analogous definition for bivariate linear operators:
\begin{definition}[list-composing bivariate linear operators]
    Let $\calG = (G_0,\dots, G_{w-1})$ and
$\calT = (T_0,\dots,T_{r-1})$ be two bivariate families of linear operators and $\calT$ be linearly-extendible.
  We say that the pair $(\calT,\calG)$ \emph{list-composes} in terms
  of $\calL$ at the set of
  evaluation points $A$ if for every linear operator $G \in \calG$ and $(x, y) \in A$, there exists a linear function $h_{G,(x, y)}: \F^s \to \F^r$
  such that for every polynomial $p \in \F[X, Y]$ we have
\[
  \calT(G(p))(x, y) = h_{G,(x, y)}(\calL(p)(x, y)).
\]
\end{definition}

Now we are ready to state the main theorem for \textit{B-LO}:
\begin{theorem}
\label{thm:main}
  If $\textit{B-LO}^A_{t, k}(\calL)$ is a \textit{B-LO} code and there exists two
  families of linear operators $\calG= (G_0,\dots, G_{w-1})$ and $\calT =
  (T_0,\dots,T_{r-1})$, $\calT$ linearly-extendible, such that
  \begin{enumerate}
  \item\label{itm:main_thm_t_distance} For some $d_1 \geq t$ and $d_2 \geq k$ such that $(d_1 - t +1)(d_2-k+1) > nr/w$, $\textit{B-LELO}_{d_1, d_2}^A(\mathcal{T})$ is a code of distance $D$.
    \item\label{itm:main_thm_list_composability} The pair $(\calT,\calG)$ list-composes in terms of $\calL$ at the set of evaluation
      points.
    \item\label{itm:main_thm_degree_preserving} $\calG$ is degree-preserving in both $X$ and $Y$.
      \item\label{itm:main_thm_diag} $\Diag(\calG) \in \F^{|\calG| \times tk}$ is the generator
        matrix of a code with distance $tk-\ell$.

    \end{enumerate}
    Then, $\textit{B-LO}^{A}_{t, k}(\calL)$ is list-decodable up to the distance $D$ with list size $q^\ell$.
  \end{theorem}

\begin{observation} 
\cref{thm:main} implies \cref{thm:intro}.

\begin{proof}
Recall the hypothesis of \cref{thm:intro}. Let $r = s - w + 1$ and let $\calG$ and $\calT$ from \cref{thm:main} be instantiated as $\calG = (L^0, ..., L^{w-1})$ and $\calT = (L^0, L^1, ..., L^{r-1})$. Then, properties 3 and 4 are automatically satisfied. For $d_1, d_2$ as in \cref{thm:intro}, property 1 is also satisfied. Finally, since for all $i \in [w]$ and $j \in [r]$ $T_j(G_i(p)) = L^{i+j}(p) = L_{i+j}(p)$, then clearly $(\calT, \calG)$ list-composes in terms of $\calL$ at $A$, satisfying property 2.
\end{proof}
\end{observation}

We also notice that the main theorem of \cite{BHKS24} is a special case of this theorem:
\begin{observation} \cref{thm:bhks-main} is a special case of \cref{thm:main}.

\begin{proof}Consider any code $LO_{k}^A(\mathcal{L})$ that satisfies \cref{thm:bhks-main}. We will show that the corresponding code $\textit{B-LO}_{k, 1}^{A'}(\mathcal{L}')$ (constructed as in \cref{sec:lo_are_blo}) then satisfies \cref{thm:main}. For $\mathcal{T}, \mathcal{G}$ from \cref{thm:bhks-main}, define $\mathcal{T}', \mathcal{G}'$ such that $T_i'(p(X, Y)) = T_i(p(X, 0))$ and $G_i'(p(X, Y)) = G_i(p(X, 0))$. Then:
\begin{itemize}
\item \cref{thm:main}-\cref{itm:main_thm_t_distance}: For $d_1 = k+\left\lfloor nr/w\right\rfloor$ and $d_2 = 1$, and $\textit{B-LELO}_{k+nr/w, 1}^{A'}(\mathcal{T}')$ is equivalent to $LELO_{k+nr/w}^A(\mathcal{T})$, which has distance $D = 1 - k/nr - 1/w$ since \textit{LELO} is MDS \cite{BHKS24}.
\item\cref{thm:main}-\cref{itm:main_thm_list_composability}: Clearly, $(\mathcal{T}', \mathcal{G}')$ list composes in terms of $\mathcal{L}'$ at $A$ since $(\mathcal{T}, \mathcal{G})$ list-composes in terms of $\mathcal{L}$ at $A$: for every $G_i' \in \mathcal{G}'$ and $a \in A$, $\calT'(G_i'(p))(a, 0) = \calT'(G_i(p(X, 0)))(a, 0) = \calT(G_i(p(X, 0)))(a) = h_{G_i, a}(\calL(p(X, 0))(a)) = h_{G_i, a}(\calL'(p)(a, 0))$ for all $p \in \F[X, Y]$.
\item\cref{thm:main}-\cref{itm:main_thm_degree_preserving}: $\mathcal{G}'$ is degree preserving in $y$ because $\text{deg}_yG_i(p) = 0$ for any $p$, and degree-preserving in $x$ since $\mathcal{G}$ is degree-preserving.
\item\cref{thm:main}-\cref{itm:main_thm_diag}: $\Diag(\mathcal{G'}) = \Diag(\mathcal{G})$, so the last condition also follows.
\end{itemize}

Therefore, if $LO_{k}^A(\mathcal{L})$ satisfies \cref{thm:bhks-main}, then there are $\mathcal{T}', \mathcal{G}'$, $d_1 = 1$, $D = 1 - k/nr - 1/w$ such that the corresponding code $\textit{B-LO}_{k, 1}^{A'}(\mathcal{L}')$ satisfies all conditions of \cref{thm:main} and gives us the list decoding bound $D = 1 - k/nr - 1/w$ -- the same  that was stated by \cref{thm:bhks-main}. Moreover, the list size is the same as well: $q^\ell$.

\end{proof}
\end{observation}

It remains to prove \cref{thm:main}.

\section{Proof of {\cref{thm:main}}}\label{sec:main_thm_proof}
In this section, we present a proof of \cref{thm:main}. This proof is a generalization of the proof for \textit{LO} codes presented in \cite{BHKS24}, which in turn abstracts the Guruswami and Wang \cite{GW12} proof for FRS codes.

Assume we get the code $\textit{B-LO}_{t, k}^A(\calL)$ that satisfies \cref{thm:main}. Let the received word be $\textbf{c} \in \left(\F^s\right)^n$. For each evaluation point $(x, y) \in A$, let $c_{(x, y)} \in \F^s$ denote the coordinate of $\textbf{c}$ that corresponds to the evaluation $\calL(p)(x, y)$. Thus, we can view the received word as $\left\{c_{(x, y)}\right\}_{(x, y) \in A}$. Below we present an algorithm to decode $\textbf{c}$ from at most $D$ errors.

\subsection{Interpolation step}
We will interpolate the polynomial $Q(X, Y, \textbf{u})$ of the following form:
\begin{equation}
Q(X, Y, \textbf{u}) = \sum_{i \in [w]}Q_i(X, Y)\cdot u_i
\end{equation}

We will require that for any $(x, y) \in A$:
\begin{equation}\label{eq:q_contraints}
\sum_{i \in [w]}\left(Q_i(M_x(X, Y), M_y(X, Y))(x, y)\cdot h_{G_i, (x, y)}\cdot c_{(x, y)}\right) = 0
\end{equation}
Here, $M_x, M_y \in \F[X, Y]^{r\times r}$ are the matrices for the linearly-extendbile family $\mathcal{T}$ (see \cref{def:b_linear_ext}). By a slight abuse of notation, $h_{G_i, (x, y)} \in \F^{r\times s}$ stands for the matrix associated with the linear transformation $h_{G_i, (x, y)}$.

For each evaluation point $(x, y)\in A$ we have $r$ homogeneous linear constraints on the coefficients of $Q$, since the equation \cref{eq:q_contraints} is $r$-dimensional. Thus, in total there are $nr$ constraints. We will require that $\text{deg}_x(Q_i) \leq d_1 - t$ and $\text{deg}_y(Q_i) \leq d_2 - k$, where $d_1, d_2$ are the degree parameters from the first condition of \cref{thm:main}. Thus, we get that the total number of free variables is:
\begin{equation}
\sum_{i\in [w]}(\deg_x(Q_i)+1)(\deg_y(Q_i)+1) = w(d_1-t+1)\left(d_2 - k + 1\right) > nr
\end{equation}
Therefore, since the number of variables is strictly greater than the number of constraints, we can indeed find a non-zero $Q$ that satisfies all constraints in polynomial time by solving the corresponding system of homogeneous linear equations.

\subsection{Close Enough Codewords Satisfy the Equation}
For each $p \in \F[X, Y]_{\substack{<t\\<k}}$, consider the following polynomial $R_p(X, Y)$:
\begin{equation}
R_p(X, Y) = \sum_{i \in [w]} Q_i(X, Y)\cdot G_i(p(X, Y))
\end{equation}

Since $G_i$ is degree preserving for both $x, y$, then we get that:
\begin{align}
&\deg_x(R_p) = \max_{i\in[w]}\left(\deg_x(Q_i)\right)+\deg_x(p) < (d_1 - t) + t = d_1 \\
&\deg_y(R_p) <  \max_{i\in[w]}\left(\deg_y(Q_i)\right)+\deg_y(p) < (d_2 - k) + k = d_2
\end{align}

For any position on which the received word $\textbf{c}$ agrees with the encoding of $p(X, Y)$, i.e. for any $(x, y)\in A$ such that $c_{(x, y)} = \mathcal{L}(p)(x, y)$:
\begin{equation}\label{eq:t(r_p)}
\begin{aligned}
\mathcal{T}(R_p) = \mathcal{T}\left(\sum_{i\in[w]}Q_i(X, Y)\cdot G_i(p)\right) &= \sum_{i\in[w]} \mathcal{T}\left(Q_i(X, Y)\cdot G_i(p)\right) = \\ &= \sum_{i\in[w]} Q_i(M_x(X, Y), M_y(X, Y))\cdot \mathcal{T}(G_i(p))
\end{aligned}
\end{equation}
The first equality of \cref{eq:t(r_p)} holds by the definition of $R_p$; the second holds by linearity of $\calT$, and the third -- by the fact that $\calT$ is linearly-extendible (see \cref{obs:b_lin_extendible}).

Then, since $(\mathcal{T}, \mathcal{G})$ list composes in terms of $\mathcal{L}$ at $A$:
\begin{equation}
\begin{aligned}
\mathcal{T}(R_p)(x, y) &= \sum_{i\in[w]} Q_i(M_x(X, Y), M_y(X, Y))(x, y)\cdot \mathcal{T}(G_i(p))(x, y) = \\ &= \sum_{i\in[w]} Q_i(M_x(X, Y), M_y(X, Y))(x, y)\cdot h_{G_i, (x, y)}\left(\mathcal{L}(p)(x, y))\right)
\end{aligned}
\end{equation}

Therefore, for any $(x, y) \in A$ such that $c_{(x, y)} = \mathcal{L}(p)(x, y)$:
\begin{equation}
\mathcal{T}(R_p)(x, y) = \sum_{i\in[w]} Q_i(M_x(X, Y), M_y(X, Y))(x,y)\cdot h_{G_i, (x, y)} \cdot c_{(x, y)} = 0
\end{equation}
The last equality holds because of the interpolation constraints on $Q$ (\cref{eq:q_contraints}). Therefore, at any agreement $(x, y) \in A$ get $\calT(R_p)(x, y) = 0$.

Notice that since $\text{deg}_x(R_p) < d_1$ and $\text{deg}_y(R_p) < d_2$, then $\left(\calT(R_p)(x, y)\right)_{(x, y)\in A}$ is the encoding of $R_p$ in $\textit{B-LELO}_{d_1, d_2}^A(\calT)$. Therefore, as according to the statement of \cref{thm:main} $\textit{B-LELO}_{d_1, d_2}^A(\calT)$ is a code of distance $D$, then if the number of non-zero evaluations in $\left(\calT(R_p)(x, y)\right)_{(x, y)\in A}$ is less than $D$, we must have $R_p \equiv 0$.

Since for each agreement point $(x, y) \in A$ such that $c_{(x, y)} = \calL(p)(x, y)$ get $\calT(R_p)(x, y) = 0$, then if there are less than $D$ errors and $p$ is the initial message polynomial, then $\left(\calT(R_p)(x, y)\right)_{(x, y)\in A}$ must have less than $D$ nonzero elemenets, meaning that $R_p \equiv 0$. Therefore, if there are less than $D$ errors, then for all possible message polynomials $p \in \F[X, Y]_{\substack{<t\\<k}}$ must have $R_p\equiv 0$. This means that our task of finding all possible message candidates reduces to finding all such polynomials $p$ for which $R_p$ is a zero polynomial.

\subsection{Solving the Equation to Recover the Codewords}
Our task is now to find all possible message polynomials $p$ for which $R_p \equiv 0$:
\begin{equation}\label{eq:main_eq_system}
Q_0(X, Y)G_0(p) + Q_1(X, Y)G_1(p) + ... + Q_{w-1}(X, Y)G_{w-1}(p) \equiv 0
\end{equation}
Clearly, given the linearity of $\mathcal{G}$, the set of all such polynomials forms a linear space. Also, given $Q$ and the description of $\mathcal{G}$, we can find this space by solving a linear system in polynomial time.

It remains to show that the space of solutions is small. First of all, let $r_1 = \max_{j\in[w]}(\deg_xQ_j)$, so we can write $Q_j(X, Y) = \sum_{i=0}^{r_1}Q_{j, i}(Y)X^i$ for all $j \in [w]$. Also, for each $0 \leq i \leq r_1$, define $r_2(i) = \max_{j\in[w]}(\deg Q_{j, r_1-i})$ -- maximum $y$-degree among of all terms in $Q_j$ with $X^{r_1-i}$. We will also write $r_2(0)$ simply as $r_2$.

Also, for each $Q_j$ let $q_{j, i, l}$ denote the coefficient of $X^iY^l$, meaning that can write $Q_j(X, Y) = \sum_{i}\sum_{l}q_{j, i, l}X^iY^l$. By the definition of $r_1$ and $r_2$, notice that $(q_{0, r_1, r_2}, q_{1, r_1, r_2}, ..., q_{w-1, r_1, r_2})$ is a nonzero vector.

For each pair of degrees $a \in [t]$, $b\in [k]$ we will associate a running index $v(a, b) = ak + b$. Then, let $p_{v(a, b)}$ be the coefficient of $X^aY^b$ in $p$, and let $\text{coeff}(p) = (p_{tk-1}, ..., p_1, p_0)$ be a vector of coefficients of $p$.
We will also define $G_{j, i}(p)\in \F[Y]$ to be the coefficient of $X^i$ in $G_j(p)$, interpreted as a polynomial in $Y$ (so $G_j(p) = \sum_{i}(G_{j, i}(p))(Y)\cdot X^i$), and let $g_{j, i, l}\in \mathbb{F}^{tk}$ be a $v(i, l)$-th row in $G_j$ ($G_j$ interpreted as an $tk\times tk$ matrix) that corresponds to the term $X^iY^l$. This means that the coefficient of $X^iY^l$ in $G_j(p)$ is $\left\langle g_{j, i, l}, \text{coeff}(p)\right\rangle$.

With these definitions, we are ready to get to the proof. First, fix any $i \in [t]$ and consider the coefficient of $X^{r_1+t-1-i}$ in \cref{eq:main_eq_system} (interpreted as a polynomial in $Y$). We get:
\begin{equation}
\sum_{j\in[w]}\left(\sum_{i'=0}^iQ_{j, (r_1-i')}(Y)\cdot G_{j, t-1-(i-i')}(p)(Y)\right)
\end{equation}

Now, for this fixed $i$, consider the coefficient of $X^{r_1+t-1-i}Y^{r_2+k-1-l}$ for some $l \in [k]$:
\begin{equation}\label{eq:coeff_xy}
\sum_{j\in [w]}\sum_{i'=0}^i\sum_{l'=0}^{r_2(i')}q_{j, (r_1-i'), (r_2(i')-l')}\cdot \left\langle g_{j, t-1-(i-i'), k-1-(l-l')+(r_2-r_2(i'))}, \text{coeff}(p)\right\rangle
\end{equation}
This expression requires some commentary. Note that $l'$ runs up to $r_2(i')$, since this is by definition the largest possible degree that could be present in $Q_{j, (r_1-i')}$. Then, for this particular $l'$, if the degree we take from $Q_{j, (r_1-i')}$ is $r_2(i') - l'$, then the degree we should take from $G_{j, t-1-(i-i')}$ in order to get the total degree of $(r_2+k-1-l)$ should be $k-1-(l-l')+(r_2-r_2(i'))$. We then use the definition of $g$ to express the coefficient of $X^{t-1-(i-i')}Y^{k-1-(l-l')+(r_2-r_2(i'))}$ in $G_j(p)$ as a corresponding inner product. This explains \cref{eq:coeff_xy}.

Now, let us figure out on which values of $\text{coeff}(p)$ \cref{eq:coeff_xy} depends. In particular, let us find the smallest value of $v(a, b)$ such that $p_{v(a, b)}$ is involved in this expression. First of all, since $\mathcal{G}$ is degree-preserving in $X$, and we are only interested in terms with $x$-degree $t-1-(i-i')$, where $i' \geq 0$, then the smallest degree of $X$ from $p$ we are interested in is $t-1-i$, which appears only in the terms with $i'=0$. Since $v(a,b) = ak+b$ first orders terms by $X$ degree, then terms with $i' > 0$ will involve only those $p_v$ with larger values of $v$. Now, if we only consider the terms with $i' = 0$ in \cref{eq:coeff_xy}, we get:
\begin{equation}
\sum_{j\in [w]}\sum_{l'=0}^{r_2}q_{j, r_1, (r_2-l')}\cdot \left\langle g_{j, t-1-i, k-1-(l-l')}, \text{coeff}(p)\right\rangle
\end{equation}
Clearly, since $\mathcal{G}$ is degree preserving in $y$, then the smallest $y$ degree of $p$ that is involved in this expression is $(k-1-l)$, so the whole expression depends only on $\{p_v: v \geq v(t-1-i, k-1-l)\}$. In particular, if we set up a system of linear equations where $v(i, l)$-th constraint equates the coefficient of $X^{r_1+t-1-i}Y^{r_2+k-1-l}$ in $R_p$ to zero, then this system would be lower-triangular, and its diagonal elements will equal to:
\begin{equation}\label{eq:diag}
\sum_{j\in[w]}q_{j, r_1, r_2}\Diag(\mathcal{G})_{j, v(t-1-i, k-1-l)}
\end{equation}
This is because in $\left\langle g_{j, t-1-i, k-1-l}, \text{coeff}(p)\right\rangle$, the coefficient of $p_{v(t-1-i, k-1-l)}$ is exactly equal to $\Diag(\mathcal{G})_{j, v(t-1-i, k-1-l)}$.

Therefore, the linear system where $v(i, l)$-th constraint equates \cref{eq:coeff_xy} to zero would be lower triangular, and its diagonal elements will be as shown in \cref{eq:diag}.

Each of these diagonal elements can be interpreted as an inner product of a nonzero vector $u = (q_{0, r_1, r_2}, q_{1, r_1, r_2}, ..., q_{w-1, r_1, r_2})$ with a vector $(\Diag(\mathcal{G})_{0, v(t-1-i,k-1-l)}, ..., \Diag(\mathcal{G})_{w-1, v(t-1-i,k-1-l)})$, which is exactly $v(t-1-i,k-1-l)$-th column of $\Diag(\mathcal{G})$. Therefore, we can get all diagonal elements of our linear system by computing $u\times \Diag(\mathcal{G})$. Since $\Diag(\mathcal{G})$ is the generator matrix of a code with distance $tk - \ell$ and $u\neq 0$, then $u\times \Diag(\mathcal{G})$ would have at most $\ell$ zeroes, meaning that in our lower triangular linear system there will be at most $\ell$ zeroes on diagonals, which in turn means that the space of solutions is at most $q^\ell$. This proves our claim.

\section{List decoding PPC}\label{sec:list_decoding_ppc}
In the original paper on permuted product codes \cite{BST24}, an algorithm similar to \cite{GW12} was used for decoding, with the main difference being that some algebraic properties of PPCs were used to limit the list size: it was done by finding the roots of a certain linearized polynomial over the well-chosen extension field, and then applying the results of \cite{Tam23} to reduce list size to constant. Below we show how \cref{thm:main} allows us to (re)prove the list-decodability of this family of codes.

As was shown in \cref{sec:ppc_blelo}, permuted product codes are \textit{B-LELO} for $\mathcal{L} = (L^0, ..., L^{s-1})$, where $L(p(X, Y)) = p(\ell_1(X), \ell_2(Y))$, and $A = \{\left(\alpha, \ell_2^{is}(\beta)\right)\}_{i = 0}^{n-1}$. 

Fix any $R$, $\epsilon > 0$. Take $s = p \approx 1/\epsilon^3$, $w \approx 1/(2\epsilon^2)$ and $x$-degree parameter $t$ such that $s-t \approx 1/\epsilon^2$.  Also let $r = s - w + 1$ (note that $r \geq t$).

Let $\ell_1(X) = X+1$ and $\ell_2(X) = \gamma X$ for some primitive $\gamma$, so $\ell_1$ indeed has order $p = s$ and $\ell_2$ has order $n = q - 1$.

Now take $\mathcal{T} = (T_0, ..., T_{r-1})$ such that $T_i = L_i$ for all $i \in [r]$, and $\mathcal{G} = (G_0, ..., G_{w-1})$ such that $G_i = L_i$ for all $i \in [w]$. Clearly, $\calT$ is linearly-extendible. Then:
\begin{itemize}
\item \cref{thm:main}-\cref{itm:main_thm_t_distance}: Let $d_1 = r \geq t$ and $d_2 = k + \left\lfloor nr/(w(r-t+1))\right\rfloor \geq k$, so indeed have $(d_1-t+1)(d_2-k+1) > nr/w$. Note that for each row of $\textit{B-LELO}_{d_1, d_2}^A(\mathcal{T})$ is an RS codeword, since the values in a fixed row $i$ are evaluations of $p$ on a set of points with the same $x$ value $\ell_1^i(\alpha)$ and distinct $y$ values $\ell_2^i(\beta), ..., \ell_2^{(n-1)s+i}(\beta)$ (these are distinct since $n$ and $s$ are coprime). Hence, each nonzero row has weight at least $n - d_2$, so each nonzero codeword of $\textit{B-LELO}_{d_1, d_2}^A(\mathcal{T})$ has at least $n - d_2$ nonzero columns, meaning that the code has relative distance $D = 1 - r/w(t-t+1) - k/n$.

\item \cref{thm:main}-\cref{itm:main_thm_list_composability}: $(\mathcal{T}, \mathcal{G})$ list-composes in terms of $\mathcal{L}$ at $A$ since $T_j(G_i(p))(a) = L_{i+j}(p)(a)$, and for any $j \in [r]$ and $i\in[w]$, $i+j \in [s]$.

\item \cref{thm:main}-\cref{itm:main_thm_degree_preserving}: $G_i(p(X, Y)) = p(\ell_1^i(X), \ell_2^i(Y))$ is degree-preserving since $\ell_1, \ell_2$ are linear operators.

\item \cref{thm:main}-\cref{itm:main_thm_diag}: Let's explore the diagonal of $G_i$. Any term $q_{ab}X^aY^b$ in $p$ becomes $q_{ab}\ell_1^i(X)^a\ell_2^i(Y)^b$ in $G_i(p)$. Given $\ell_1, \ell_2$ as we chose, initial term $X^aY^b$ contributes to the new such term the coefficient $q_{ab}\gamma^{ib}$. Therefore, $\text{Diag}(\mathcal{G})_{i, ak + b} = \gamma^{ib}$, if index $ak + b$ corresponds to coefficient $X^aY^b$. Hence, $\Diag(\mathcal{G})$ is formed by stacking $t$ generator matrices of $RS[w, k]$, meaning that its distance is $(k-w+1)t = tk - (w-1)t$.
\end{itemize}

Therefore, by \cref{thm:main}, this code is list decodable from the distance $1 - r/w(r-t+1) - k/n$ with list size $q^{(w-1)t}$. Note that:
\[
\frac{r}{w(r-t+1)} < \frac{s}{w(s-t-w)} \approx \frac{1/\epsilon^3}{1/2\cdot1/\epsilon^2\cdot 1/2\cdot 1/\epsilon^2} = O(\epsilon)
\]
\[
\frac{k}{n} = \frac{tk}{sn}\cdot \frac{s}{t} = R\cdot \frac{s}{t}\approx R\cdot \frac{1/\epsilon^3}{1/\epsilon^3 - 1/\epsilon^2} = R\cdot \frac{1}{1-\epsilon}\leq R(1+2\epsilon) = R+O(\epsilon)
\]
\[
(w-1)t < wt \approx \frac{1}{2\epsilon^2}\cdot \left(\frac{1}{\epsilon^3} - \frac{1}{\epsilon^2}\right) < \frac{1}{\epsilon^5}
\]

Therefore, PPC is list decodable from the distance $1 - R - O(\epsilon)$ with list size $q^{1/\epsilon^5}$. 

\section{Acknowledgments}

The authors would like to thank Madhu Sudan for helpful conversations and guiding directions during this project. 

\section{Open questions}\label{sec:open_questions}

Below we suggest some open questions for further research:
\begin{itemize}
\item \textbf{New \textit{B-LO} codes.} Since \textit{B-LO} codes are significantly more general than \textit{LO} codes, there might be previously unknown list decodable codes that are instances of \textit{B-LO}.
\item \textbf{More variables.} The definitions suggested in this work might be extended beyond bivariate linear operators. It might be worth exploring a more general case of $L_i: \F[X_1, ..., X_n]\to\F[X_1, ..., X_m]$ linear operators and see if this yields any useful list decodable codes.
\end{itemize}

\bibliographystyle{alpha}
\bibliography{ref}

\end{document}